\documentclass[journal]{IEEEtran}

\ifCLASSINFOpdf

\else

\fi

\usepackage{cite}
\usepackage{amsmath,amssymb,amsfonts}
\usepackage{algorithmic}
\usepackage{graphicx}
\usepackage{textcomp}
\usepackage{color}
\usepackage{soul}

\begin{document}

\title{Multimodal Estimation of End Point Force During Quasi-dynamic and Dynamic  Muscle Contractions Using Deep Learning}

\author{Gelareh Hajian,~\IEEEmembership{ Member, IEEE}, Evelyn Morin,~\IEEEmembership{Senior Member, IEEE}, and Ali Etemad,~\IEEEmembership{Senior Member, IEEE}
\thanks{All authors are with the Department of Electrical and Computer Engineering, Queen's University, Kingston, ON, K7L 3N6, Canada (e-mail: \{14gh6, evelyn.morin, ali.etemad\}@queensu.ca). }}

\maketitle

\begin{abstract}

Accurate force/torque estimation is essential for applications such as powered exoskeletons, robotics, and rehabilitation. However, force/torque estimation under dynamic conditions is a challenging due to changing joint angles, force levels, muscle lengths, and movement speeds. We propose a novel method to accurately model the generated force under isotonic, isokinetic (quasi-dynamic), and fully dynamic conditions.  Our solution uses a deep multimodal CNN to learn from multimodal EMG-IMU data and estimate the generated force for elbow flexion and extension, for both intra- and inter-subject schemes. 
The proposed deep multimodal CNN  extracts representations from EMG (in time and frequency domains) and IMU (in time domain) and aggregates them to obtain an effective embedding for force estimation.
We describe a new dataset containing EMG, IMU, and output force data, collected under a number of different experimental conditions, and use this dataset to evaluate our proposed method. The results show the robustness of our approach in comparison to other baseline methods as well as those in the literature, in different experimental setups and validation schemes. The obtained $R^2$ values are 0.91$\pm$0.034, 0.87$\pm$0.041, and 0.81$\pm$0.037 for the intra-subject and 0.81$\pm$0.048, 0.64$\pm$0.037, and 0.59$\pm$0.042 for the  inter-subject scheme, during isotonic, isokinetic, and dynamic contractions, respectively. Additionally, our results indicate that force estimation improves significantly when the kinematic information (IMU data) is included. Average improvements of  13.95\%, 118.18\%, and 50.0\%  (intra-subject) and 28.98\%, 41.18\%, and 137.93\% (inter-subject) for isotonic, isokinetic, and dynamic contractions respectively  are achieved.

\end{abstract}


\section{Introduction}
\label{sec:introduction}

Accurate estimation of muscle end-point force/torque is necessary for a variety of applications such as powered exoskeletons, robotics, and rehabilitation systems. Since the demonstration of the relationship between muscle electrical activity and generated force in humans \cite{inman1952}, the surface electromyogram (EMG) has been widely used as a non-invasive estimator of the generated force and joint torque. EMG signals can be used to estimate the neural command for muscle contraction and force generation in motor control \cite{motorcontrol}, and for designing control algorithms based on the user’s intention for robotic arm \cite{robotsapplication,su20033}, powered exoskeleton \cite{power_exoskeleton}, and prosthesis control \cite{castellini2009,su2007towards,GelarehMoveEst2022} applications. In ergonomics \cite{ergonomics} and clinical biomechanics \cite{biomechanics2009} EMG can be used to better estimate joint loading and muscle tension to prevent musculoskeletal injuries. And it can contribute in assistive devices \cite{hal2010} by enabling feedback control to provide assistance in tasks such as lifting.

Classical approaches to EMG-based force modeling have used musculoskeletal models \cite{musculoskeletalmodel} and Hill’s muscle model \cite{hill1938heat}, which require precise estimates of physiological parameters along with a number of assumptions, as some parameters involved in the models cannot be measured experimentally \cite{neuromusculoskeletalmodelling}. The level of accuracy in these methods highly depends on the physiological parameters which are often assumed to be constant among subjects as a means of simplification. 
Thus, these models are generally limited in application and subject to high variability given the experiment setups and factors relating to the models and subjects.

Another category of solutions focuses on data-driven approaches in which a model aims to estimate output force (or torque) from input EMG data without requiring knowledge of the muscle, joint dynamics, or the subject. Examples include system identification techniques such as polynomial estimation \cite{clacny2017}, fast orthogonal search (FOS) \cite{korenberg1989FOS,hajian2020channelselection}, and parallel cascade identification (PCI) \cite{hashemi2015PCI}. More recently, machine learning solutions have been used to \textit{learn} mappings between EMG and force. Examples include the use of artificial neural networks (ANN) \cite{mobasser2012ANN,hajian2019ANN,featureselectionforce2016} and Support Vector Machine (SVM) \cite{featureselectionforce2016, svrziai2011,castellini2009}, which often highly depend on the quality and variety of hand-crafted features extracted and provided to the model. \textit{Deep learning} solutions, on the other hand, are capable of automatically learning the features necessary to maximize performance and can generalize better across subjects. The feasibility of using deep learning, such as convolutional neural networks (CNN) and long-short term memory (LSTM) networks for developing a generalized solution has been investigated \cite{CNNLSTM}, where the results have shown the potential of deep learning approaches for modelling real-time, subject-independent situations.

Most existing work on EMG-based force estimation has focused primarily on EMG amplitude estimate as the only feature, whether through hand-crafted features \cite{hajian2020channelselection,hashemi2015PCI,mobasser2012ANN} or learned representations (deep learning methods) \cite{CNNLSTM}. 
However, it has been shown that a considerable amount of valuable information regarding muscle activation is present in the EMG frequency domain \cite{farinafreqinter,meanfreqwaveletForce}. As a result, methods for estimating force that incorporate both time and frequency domain representations of EMG has a potential to achieve comprehensive solutions for EMG-based force estimation \cite{Gelareh2021generalizedcnnstatic}.

Despite the fact that most muscle force generation happens during dynamic contractions, most existing force prediction models utilize static or quasi-static experimental setups. During dynamic contractions, factors such as changes in joint angle and muscle length, muscle shape variations during movement, and contraction velocity \cite{staudenmann2010}, complicate the EMG-force relationship. Past studies have attempted to estimate force under such conditions \cite{hashemi2015PCI,liu1999dynamic,luo2019NNForceEst,mobasser2012ANN}. Hashemi et al. \cite{hashemi2015PCI} used PCI modeling, where angle-based EMG amplitude calibration was used to minimize the effects of joint angle related factors. Although the proposed calibration method improved the outcome, additional data collection was needed to obtain the calibration values \cite{hashemi2015PCI}. 

Neural-network-based algorithms and SVM models have also been employed to capture the non-linearity and dynamics of the EMG-force relation \cite{liu1999dynamic,luo2019NNForceEst,meanfreqwaveletForce,mobasser2012ANN}. Luo et al. utilized three domains fuzzy wavelet neural network to estimate the  force,  where the EMG amplitude was used as the input to the model \cite{luo2019NNForceEst}. Force and EMG signals were recorded from a variety of hand grip scenarios from 2 subjects \cite{luo2019NNForceEst}.
Bai et al. used a two-layer ANN for intra-subject force modelling, where the EMG mean frequency acquired by continuous wavelet transform was used to derive the EMG-force relation under dynamic contractions \cite{meanfreqwaveletForce}. Their results suggested that using frequency domain information can yield promising force estimation results. Zhang et al. extracted six time domain features from EMG, and used three machine learning methods, ANN, SVM, and locally weighted regression, to estimate the grasping force \cite{featureselectionforce2016}, achieving the best result using the ANN. In other studies, SVM has also been utilized for estimating force during either flexion and extension, or grasping force \cite{svrziai2011, svrGripforceEST,castellini2009}. SVM outperformed ANN in one case \cite{castellini2009}.

Capturing and utilizing the motion associated with dynamic force generation, for example in the form of kinematic data \cite{staudenmann2010,countermovement}, can provide a rich source of information that can contribute to estimation of the generated force during dynamic conditions. Motion data has been used along with EMG for force estimation using ANN \cite{mobasser2012ANN,luh1999isokineticANN}. Mobasser et al. used ANN for intra-subject force estimation, under isometric, isotonic, and light load dynamic conditions \cite{mobasser2012ANN}. The models used the linear envelope of EMG signals along with the elbow joint angle, angular velocity, and acceleration as inputs. Their method predicted the nonlinear relation between the EMG, angle and velocity, and the force generated at the wrist. The predictions improved when acceleration data were added to the model. Luh et al. also achieved good performance in elbow joint torque estimation under isokinetic conditions with EMG amplitude, joint angle, and joint angular velocity as inputs to an ANN \cite{luh1999isokineticANN}.

With the widespread popularization of wearable devices, Inertial Measurement Units (IMU) (which often contain an accelerometer, a gyroscope, and a magnetometer) have become the standard solution for low-cost motion monitoring. Oboe et al. \cite{emgIMU} used the IMU and four EMG signals as inputs to ANN for estimating four different weights held in the hand, during a standardized movement defined for their experiment. Since their model did not accurately estimate the weight lifted, they suggested that the model could be used for binary classification, i.e. to classify the light versus heavy weights.

In this paper we tackle the problem of end point force estimation during dynamic contractions. We target elbow flexion and extension using high density (HD) surface EMG recorded from the flexor and extensor muscles, since using multiple EMG channels can improve the performance of force estimation in comparison with single bipolar electrodes due to the enhanced neuromuscular information \cite{clacny2017,Gelare2018FOS}. EMG, motion data (obtained from a wearable IMU device mounted on the forearm) and ground-truth force data were collected under quasi-dynamic (controlled force/or controlled velocity) and dynamic (no control on force and velocity) conditions. To develop a robust, generalized, and end-to-end solution, we propose the use of deep learning, in particular, a deep multimodal CNN to learn EMG in both time and frequency domains as well as motion signals in time domain. The results of our experiments show the robustness of our method in comparison to other baseline methods as well as those in the literature. We also performed ablation experiments to validate the added value of each modality and we explored several fusion strategies to further validate our approach. Our contributions in this paper can be summarized as follows.
(\textbf{1}) We propose a new solution based on a CNN pipeline for force estimation during quasi-dynamic and dynamic contractions, exploiting both time and frequency domain information of EMG signals to improve the performance. The model is evaluated for \textit{intra-subject} and the more challenging \textit{inter-subject} force estimation. To the best of our knowledge, our work is the first study to develop a generalized model for force estimation under dynamic conditions and in both intra- and inter-subject settings. Our solution obtains accurate results in both evaluation schemes, outperforming other methods in the field and setting new state-of-the-art values, demonstrating the effectiveness of our method. 
(\textbf{2}) We utilize both EMG and IMU data for generated force estimation and our results show that incorporating IMU data considerably contributes to the accuracy of the model.
(\textbf{3}) We collect a new and comprehensive dataset that contains EMG, IMU, and exerted force under a variety of operational conditions.

\section{Methodology}
\label{sec:methodology}

\subsection{Experimental Setup and Data Collection}

The experiments were conducted in the Queen’s University Human Mobility Research Lab, at Hotel Dieu Hospital, Kingston, Canada. Sixteen healthy participants, 8 females and 8 males, with an average age of 26$\pm$9 years, were recruited for this study. The experimental procedures were approved by the Health Sciences and Affiliated Teaching Hospitals Research Ethics Board (HSREB) of Queen’s University. Participants provided informed consent before participating in the experiment. 

\begin{figure}
    \centering
    \includegraphics[width=1\columnwidth]{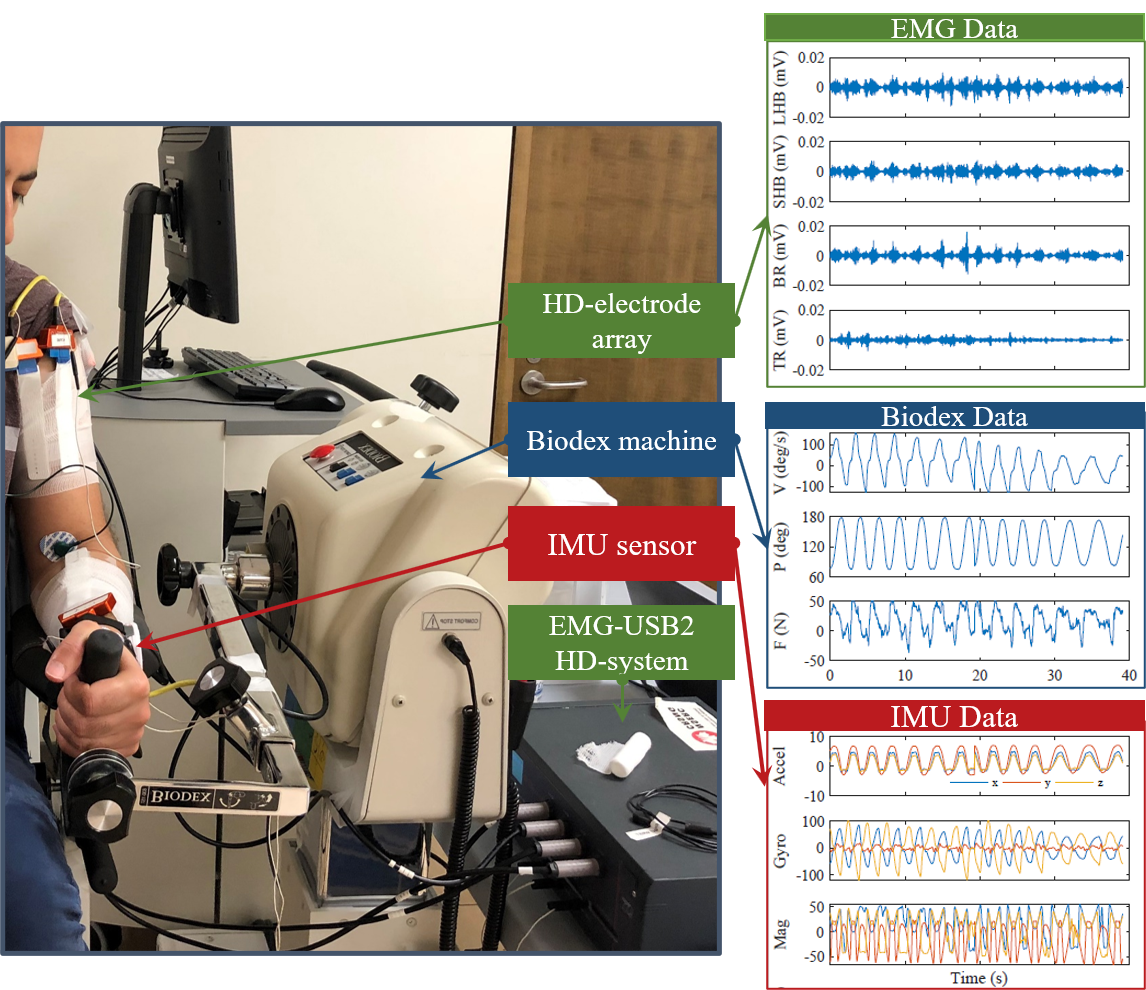}
    \caption{The experimental setup with a sample data recorded from one subject during dynamic condition are shown. EMG signals recorded from; long head of the biceps brachii (LHB), short head of the biceps brachii (SHB), brachioradialis (BR), and triceps brachii (TR). Motion data recorded by the IMU sensor; acceleration (Accel),  gyroscope (Gyro), Magnetometer (Mag). Data recorded by the Biodex; velocity (V), position (P), and Force (F).}
    \label{experimental_setup}
\end{figure}

The Biodex (model 840-000) \cite{biodex}, a multi-joint device which can be used to study the human musculoskeletal system, was used to control the motion of the arm, while measuring the elbow joint angle, speed of movement, and generated torque. The Biodex was set up for elbow flexion and extension of the right arm, as shown in Fig. \ref{experimental_setup}.
The experimental protocol included three specific dynamic motions, isotonic-nonisokinetic, isokinetic-non-isotonic, and fully dynamic, for which the Biodex isotonic mode, passive mode, and isokinetic mode were used respectively.  Participants were seated and the arm was held in place at the elbow with a cushion under the elbow (for comfort and to stabilize the elbow), while they were instructed to hold the Biodex handle (specific Biodex attachment for elbow flexion/extension). Before starting the experiment, the elbow range of motion (ROM) for flexion and extension were recorded. Flexion ROM was defined as the range from a $90$ degree elbow joint angle to the point of maximal flexion. Extension ROM was defined as the range from maximal flexion to the $90$ degree joint angle. The ROM was re-recorded at the beginning of each experimental condition with the goal of keeping it constant during the experiment for each individual. The ROM was different among subjects, based on the bulk of the upper arm which affects how much a participant could flex the elbow. The ROM among subjects had a range of $74$ to $86$ degrees.

Participants performed a series of flexions followed by extensions, where we used the Biodex to define the number of contractions. For the isotonic contraction, the dynamometer requires the subject to meet a minimum selected torque limit in order to move the input arm. Therefore, the speed of movement is variable whereas the torque is constant. Isotonic contractions were performed first for flexion and then for extension, for $3$ torque levels, $5$, $8$, and $12$ Nm applied to the elbow joint, where the average of the wrist forces were 12.53$\pm$0.49, 20.44$\pm$0.79, and 30.67$\pm$1.19 N. 
In this experiment, the range of force levels applied by the Biodex was sufficient to acquire low to high force contractions, since the applied force levels were between 15\% to 60\% maximum isometric voluntary contractions (MVC) for all subjects. MVCs were recorded for elbow flexion and extension at $90$ degree elbow joint angle. 
During this condition there was no constraint on the movement velocity. For the isokinetic condition, there were $3$ velocity levels: $60$, $90$, and $180$ deg/s. No minimum torque level was required for this experiment. During the fully dynamic condition, there were no limitations on the applied torque level and velocity during the movement, and the subjects moved their arm freely with different velocity and torque levels. For each subject, the data were collected in one session and $12$ trials per condition ($2$ sets of $6$ continuous repetitions with $30$ seconds rest between sets). Appropriate rest periods - at least $10$ minutes and more if needed - were provided between conditions to avoid muscle fatigue. For dynamic contraction, $3$ sets were performed, which resulted in a total of $36$ trails. 

The exerted force was applied to an attached bar connected to the Biodex as shown in Fig. \ref{experimental_setup}. The length of this bar was adjusted for each participant, and was recorded to convert the generated torque at the elbow to force values at the wrist. All the data recorded by the Biodex were sampled at $1250$ Hz.


The EMG data were collected using the EMG-USB2 HD-system \cite{BIOELECTRICA}, operated in a referenced mono-polar mode. Prior to electrode placement, the skin was shaved if required, cleaned, and abraded using an abrasive conductive gel. EMG sensor arrays were attached to the skin using adhesive pads with wells filled with conductive paste over the electrode contacts. EMG signals were recorded from the long head and short head of the biceps brachii, the brachioradialis, and the triceps brachii, using $4$ linear HD-electrode arrays with $8$ monopolar channels ($5$ mm spacing). For the biceps, the fourth electrode of each array was placed at the recommended SENIAM location \cite{sensorlocation}. For the brachioradialis, the fourth electrode was placed at one-third the length of the forearm measured from the elbow. For the long head of the triceps brachii, electrodes were placed at 50\% of the distance between the posterior crista of the acromion and the olecranon at $2$ finger widths medial to the line between them. Each electrode array was connected to the EMG-USB2 via an adapter, where each adapter had its own reference electrodes. Standard ECG pre-gelled electrodes with Ag/AgCl contact were used as reference electrodes, which were placed on regions with lower myoelectric activity. For the brachioradialis, the reference electrode was located on the wrist, while for the long head and short head of the biceps and for the triceps brachii they were placed on the elbow and fossa cubit (tendon). A driven right leg (DRL) circuit was used to reduce the $60$ Hz interference by attaching two reference electrodes on the right and left wrists. EMG signals were recorded with a sampling frequency of $2048$ Hz, and were hardware band-pass filtered with cut-off frequencies of $10$ and $500$ Hz.

To track and record the movement of the arm, a Shimmer wearable IMU sensor \cite{shimmer} was placed on the back of the forearm, $4$ cm from the ground electrode's location on the wrist. This location was chosen as it showed less muscle movement during the experiment, which reduced recording noise due to unwanted movement of the IMU.

The IMU has three sensors namely a triaxial accelerometer, a triaxial gyroscope, and a triaxial  magnetometer, all of which were recorded at a $500$ Hz sampling rate. The experimental setup, showing a subject seated in the Biodex machine, the EMG-USB2 HD-system, the HD-electrodes, and IMU sensor are shown in Fig. \ref{experimental_setup}.

\subsection{Data Pre-processing}
Torque, position, and velocity signals recorded from the Biodex, sampled at $1250$ Hz, were up-sampled using linear interpolation to $2048$ Hz, in order to match the sampling frequency of the EMG. The IMU data were similarly up-sampled from $500$ to $2048$ Hz. Then, differential HD-EMG signals were obtained by subtracting neighboring channels, resulting in $7$ differential channels from each array. Each differential channel was further band-pass filtered with cut-off frequencies of $10$ Hz and $500$ Hz using an eighth-order Butterworth filter. The Biodex data were smoothed using a $300$-point moving average filter. The IMU data were low-pass filtered using a Savitzky-Golay filter, with a window length of $400$ points. Prior to processing the data obtained during contractions, the rest periods between sets were manually flagged and discarded.

Given the stochastic nature of EMG, it is common to segment EMG signals into short-duration blocks where wide-sense stationarity holds \cite{ameri2019cnn,farrell2007optimal}. Accordingly, windows of EMG data ($100$-$200$ ms) are usually exploited for pattern recognition applications \cite{ameri2019cnn,farrell2007optimal}. Nielsen et al. investigated the sensitivity of force estimation performance with respect to the duration of the processing window size under isometric conditions \cite{forceESTFarina2011}, and the results did not differ significantly for a segment length of $100$ ms or longer. In this paper, we considered segment lengths of $50$ ms, $100$ ms, and $150$ ms with an overlap of half the segment length. Our analysis, as we will demonstrate later in Section \ref{windowsize}, showed that $50$ ms windows performed the best. Therefore the rest of the analysis is based on $50$ ms segments.

\subsection{Proposed Method}
\label{proposed_cnn_ens}

The force generated by muscles can be characterized with both the EMG from which it is produced, as well as the resulting motion. Thus, in order to take full advantage of the information available in our experimental data, we use a multimodal approach that uses both EMG and IMU. Moreover, as EMG often contains valuable information in both time (EMG$_{time}$) and frequency domains (EMG$_{freq}$), we explore both domains through our pipeline.

\begin{figure*}
     \centering
    \includegraphics[width=0.80\textwidth]{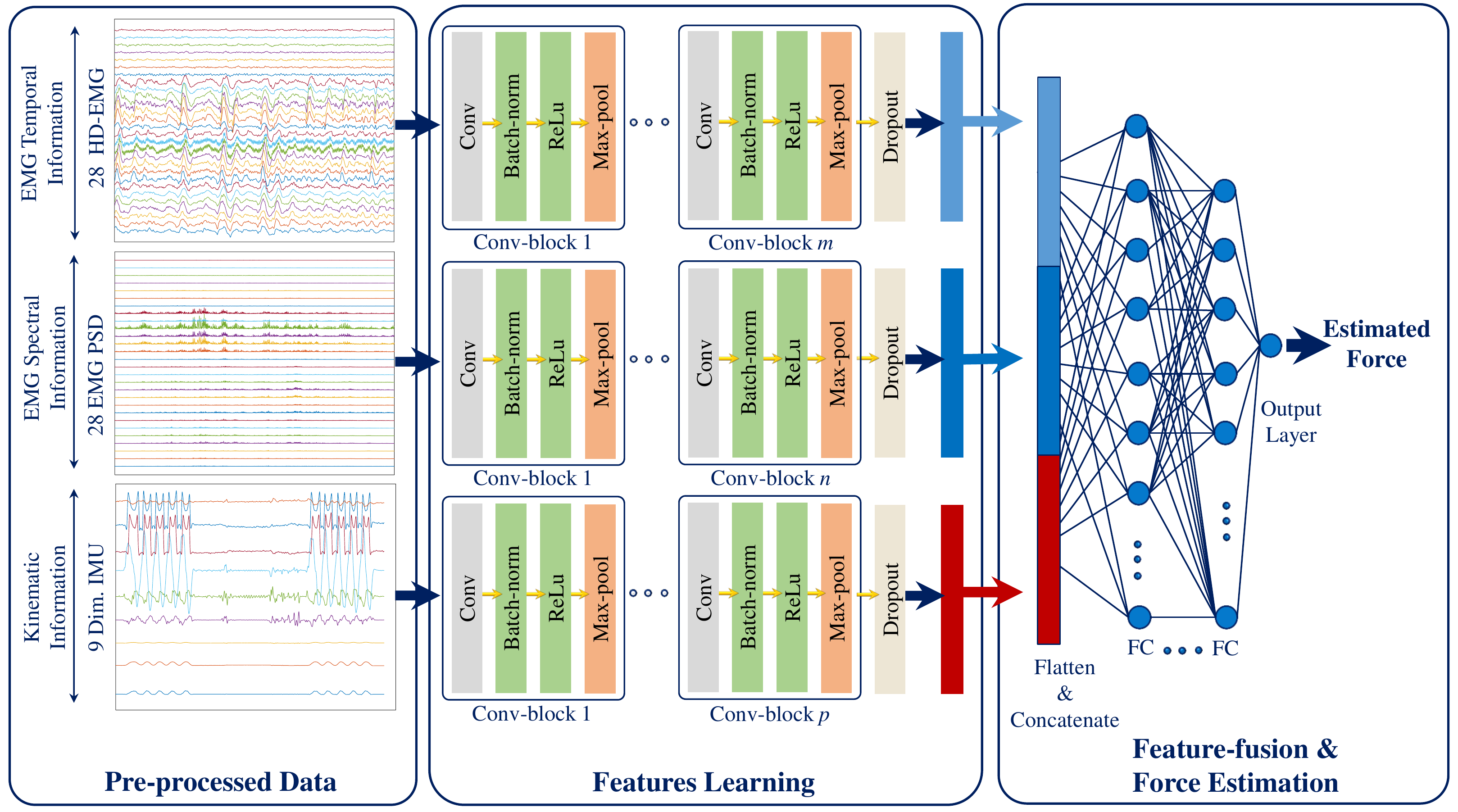}
    \caption{An overview of the proposed deep multimodal CNN.}
    \label{deep_cnn_ensemble}
\end{figure*}

An overview of the proposed deep multimodal CNN framework for force modelling is shown in Fig. \ref{deep_cnn_ensemble}. First, EMG$_{freq}$ is calculated based on the power spectral density (PSD) using the periodogram method applied on each EMG segment from all channels. Then, the pre-processed and segmented EMG$_{time}$, EMG$_{freq}$, and the time domain IMU data are used as inputs to separate CNNs, called base learners. The base learners are described in the following subsections. Each base learner is trained separately to extract the necessary features from its respective input (feature learning block). Next, in the feature fusion block, the learned features are fused, obtaining a multimodal feature map of EMG$_{time}$, EMG$_{freq}$, and IMU. This is then followed by fully connected (dense) layers which act as a shallow neural network to weight the obtained features and a regression layer to estimate the induced force at the wrist.

\subsubsection{Base Learners for EMG and IMU Data}
\label{sec:Base Learners}

CNNs are extensions of standard neural networks. While originally proposed for analysis of image and video data, today they are used in a variety of problem domains related to biological signals \cite{ameri2019cnn}. CNNs are capable of dealing with high-dimensional raw data, with no need for feature extraction, as they learn from the training data to extract the necessary features. CNNs are generally made up of several types of layers. The proposed base learner CNN architecture (shown in Fig. \ref{deep_cnn_ensemble}) comprises a number of blocks (conv-blocks), each block consisting of a convolutional layer, followed by a batch normalization and a nonlinear activation function (rectified linear unit (ReLU) is used here), and lastly a max-pooling layer.

In the first convolutional layer, the input is convolved with a set of $R$ kernels (filters), followed by the addition of the bias terms. Next, the output will go through the activation function, where a nonlinearity will be added and feature maps are produced by:

\begin{equation}\label{conv}
    \bold {FM}_r=\it {g}({\sum_{r}(\bold {X}_i*\bold{K}_r+\bold{b}_r})),
\end{equation}
where
\begin{equation}\label{relu}
  g(x)=max(x,0).
\end{equation}

$\bold {FM}_r$ denotes the produced feature maps after the ReLU activation, $\bold {X}_i$ is the input to the layer, $\bold {K}_r$ is the $r^{th}$ convolution kernel (with dimension $m \times m$), $\bold {b}_r$ is the bias term for $r = 1,...,R$, and the asterisk (*) denotes the convolution operation. A convolutional layer is usually followed by a pooling layer, a non-learnable layer, which performs sub-sampling on the feature maps. Maximum or average pooling is usually applied to summarize the activation outputs within a rectangular neighborhood with a maximum or an averaged value, to reduce the dimensionality of the feature map and decrease computation, and to help avoid overfitting. 
Fully connected (FC) layers  are often used after several convolution and pooling layers, where each neuron of the FC layers receives input from all the neurons of the previous layer. The final layer is the loss layer which determines how training is performed by minimizing the error between the predicted and true values, for classification or regression problems.

As our goal in this study is to perform both intra- and inter-subject modeling, different numbers of conv-blocks were used for these two settings, for EMG$_{time}$, EMG$_{freq}$, and IMU, denoted by \textit{m}, \textit{n}, and \textit{p} respectively.

To avoid overfitting, \textit{L}2 regularization was used. Additionally dropout was performed after the final conv-block. For the EMG$_{time}$ learner, inputs consisted of segmented pre-processed EMG data ($28$ differential signals), while the inputs to the EMG$_{freq}$ learner were the PSDs of the $28$ EMG channels. Both EMG learners were designed with the same number of conv-blocks. The input layer for the IMU learner was fed with the segmented and pre-processed IMU data recorded by the triaxial accelerametor, gyroscope, and magnetometer ($9$ channels in total). The size of the input layers of the base learners for EMG$_{time}$, EMG$_{freq}$, and IMU were (102$\times$28), (51$\times$28), and (102$\times$9) respectively for $50$ ms segment lengths, (204$\times$28), (102$\times$28), and (204$\times$9) for $100$ ms segment lengths, and (307$\times$28), (153$\times$28), and (307$\times$9) for $150$ ms segment lengths. Since the EMG$_{time}$, EMG$_{freq}$, and IMU signals are used as inputs to separate base learner CNNs, there is no need for them to be the same size. In the following subsections, we describe the details of the networks for each of the two schemes (intra- and inter-subject) considered in this study.

\subsubsection{Intra-subject Modeling}\label{convblocksIntra}

A number of hyper-parameters for each of the base learners was explored and tuned to achieve the best results. The tuned hyper-parameters include: the number of conv-blocks, number of filters and their sizes for each convolutional layer, 
batch sizes, number of training epochs, and dropout rates. The optimum values for these parameters for each learner are presented in Table \ref{Table-Hyperintra-inter} for intra-subject force modelling. In order to avoid over fitting in this case we used simple architectures for each base learner. During the hyper-parameter tuning process, the performance of both training and validation sets are considered to ensure that their $R^2$ values are not diverging (as shown in Fig. \ref{inter_base_Learner_fc}). Additionally, Maxpooling and dropout layers are used for both schemes to further reduce the risk of overfitting.

The batch size for this scheme was set to $256$ since sizes below that threshold resulted in longer training times, while not improving the performance, and batch sizes larger than that decreased the regression accuracy. For the number of epochs $100$ was selected because higher numbers did not improve the performance, and resulted in longer training times, while fewer epochs reduced the performance. The selected dropout rate for all base learners was $0.5$. The obtained parameters were used for all experimental conditions (isotonic, isokinetic, and dynamic contractions).

\begin{table}[!]
\caption{Hyper-parameters (number of convolution blocks, number of filters, filter sizes, and Maxpool sizes) for all experimental conditions and schemes.  
}
\label{Table-Hyperintra-inter}
\centering
\scalebox{0.80}{
\begin{tabular}{lcccccc}
\hline
 &  No. Conv. & No. Filters & Filter Size & Maxpool \\ \hline\hline
\textbf{Intra-subject Scheme} \\ \hline
EMG$_{time}$          & $m=2$             & $16$, $16$          & $3 \times 3$  & $3 \times 3$  \\ 
EMG$_{freq}$          & $n=2$             & $16$, $16$          & $3 \times 3$ & $3 \times 3$  \\ 
IMU               & $p=2$             & $32$, $64$          & $2 \times 2$   & $2 \times 2$ \\ 

\hline\hline
\textbf{Inter-subject Scheme} \\\hline
\textit{Isotonic and Isokinetic Conditions}   \\ 
EMG$_{time}$          & $m=2$           & $64$, $128$         & $3 \times 3$                  & $3 \times 3$     \\ 
EMG$_{freq}$          & $n=2$            & $64$, $128$          & $3 \times 3$                 & $3 \times 3$     \\ 
IMU               & $p=2$            & $64$, $128$         & $2 \times 2$                  & $2 \times 2$     \\ 
\hline
\textit{Dynamic Condition}   \\ 
EMG$_{time}$         & $m=2$           & $64$, $128$         & $3 \times 3$                      & $3 \times 3$      \\ 
EMG$_{freq}$          & $n=2$            & $64$, $128$         & $3 \times 3$                      & $3 \times 3$      \\ 
IMU               & $p=3$            & $64$, $128$, $128$     & $2 \times 2$                       & $2 \times 2$     \\ 
\hline

\end{tabular}}
\end{table}

\subsubsection{Inter-subject Modeling} \label{convblocksInter}
Given that inter-subject modeling of force is a considerably more complex problem given the physiological differences across subjects, and the need to generalize one specific model to all the subjects in the dataset, deeper networks are required in order to model the non-linearities within the problem space. Therefore, we expanded our search for the optimum hyper-parameters and explored deeper networks for the base learners. The same set of parameters was found
to provide the best results for the three conditions (isotonic, isokinetic, and dynamic), with the exception of the inter-subject dynamic condition where a deeper model was needed, as shown in Table \ref{Table-Hyperintra-inter}.

It was interestingly observed that the batch normalization operation had a considerable negative impact on the performance for inter-subject modeling. This is due to the fact that since each batch may contain EMG segments from different subjects, normalizing the batch implies that a particular subject's EMG might be normalized differently based on the batch, resulting in major discrepancies among different segments of EMG from the same subject. As a result, we removed batch normalization from the conv-blocks.
The batch size was set to $512$ and $230$ epochs were used for training. Similar to intra-subject modeling, the selected dropout rate for all base learners was $0.5$.

\subsubsection{Feature-Level Fusion}

Successive to extraction of effective features from each input data type (EMG$_{time}$, EMG$_{freq}$, and IMU) by the base learners, a fusion strategy is required to aggregate the information with the goal of estimating the generated force. To this end, all extracted features are concatenated to generate a single feature-set to be used by the output force estimation layer. This strategy, which we call \textit{feature-level fusion}, is presented in the last block of Fig. \ref{deep_cnn_ensemble}. Feature-level fusion enables the extraction of required features from each individual input modality, followed by the use of a single model for estimating force based on all the extracted features. This approach allows the force estimator to take into account all the available information at once to exploit complimentary information in the respective feature spaces. Feature-level fusion was compared with two alternative fusion methods: \textit{input-level fusion} and \textit{score-level fusion}. Input-level fusion requires a single model to learn to extract effective features from the concatenated raw inputs. Score-level fusion requires that force be estimatable based on each of the modalities, followed by the averaging step which only serves to reduce the variance in the output.

\subsubsection{Output Force Estimation}

In the proposed deep multimodal CNN architecture, shown in Fig. \ref{deep_cnn_ensemble}, after feature learning block, a flatten layer transforms the two-dimensional matrix of features obtained from the CNN base models into a vector. The flatten-concatenated set of extracted features are fed into the FC layers,  a shallow neural network to generate a force prediction. The number of FC layers and their neurons for each modelling scheme, under different experimental conditions are given in Table \ref{Table-Hyperinter_fc}.

\begin{table}[t]
\caption{Number of FC layers and associated neurons for intra- and inter-subject force modelling.
}
\label{Table-Hyperinter_fc}
\centering
\scalebox{0.85}{
\begin{tabular}{lcc|cc}\hline
          & \multicolumn{2}{c}{Intra-subject} & \multicolumn{2}{c}{Inter-subject} \\\hline
          & No. FC       & FC Size      & No. FC    & FC Size         \\\hline\hline
Isotonic   & $1$  & $128$    & $2$   & $128$, $256$        \\
Isokinetic & $1$  & $128$   & $2$    & $128$, $128$         \\
Dynamic    & $1$   & $128$    & $3$    & $128$, $256$, $256$  \\\hline
\end{tabular}}
\end{table}

\subsection{Implementation and Training}
We implement our proposed architecture and all the analysis, using Keras with a TensorFlow backend, on an Nvidia GTX 2080 Ti GPU. The models for both intra- and inter-subject schemes were trained with the adaptive moment estimation (ADAM) algorithm as the optimizer to update the network weights during back-propagation since this optimizer has been proven to be more efficient in computing the stochastic gradient problem \cite{adamoptimizer}. Learning rate (l$_{r}$) of $0.001$ was used with exponential decay rates for the first and second movement estimates of $\beta1=0.9$, and $\beta2=0.999$ respectively.

For both intra- and inter-subject evaluation schemes, the dataset was randomly split into train, validation, and holdout test sets. First, $10\%$ of the data was assigned as a holdout set, and $5$-fold cross-validation was used on the remaining train and validation sets. Hyper-parameters such as parameters for the base learners, including the type and number of layers, as well as the width of the fully connected layers in the pipeline, and the fusion strategy, were all tuned based on the cross-validation strategy. The holdout test set, which contains recordings completely unseen by the model was used to evaluate the performance of our model, using the set of best hyper-parameters.

\subsection{Evaluation Metric}

To evaluate the performance of our proposed model, R-squared ($R^2$) was used, which is calculated as
  follows:
  \begin{equation}\label{r2}
    R^2=1-\frac{\sum_{i=0}^N(F^{Est}_i-F_i)^2}{{{\sum_{i=0}^N}(F_i-\overline{F_i})^2}}, 
  \end{equation}
where $N$ is the number of data samples, $F_i$ is the $i$th measured force sample, $F^{Est}_i$ is the corresponding estimated force, and $\overline{F_i}$ is the average of $F_i$. The numerator in the second term of the equation is the total mean squared error (MSE) of the estimates, while the denominator is the total variance of the force.

\subsection{Evaluation Experiments} \label{evaluation}

First, in order to better evaluate the contribution of our work, we compare the performance of our deep multimodal CNN approach to other works in the area. We re-implemented several other published methods to apply on our data for this purpose. The study by Mobasser et al. \cite{mobasser2012ANN} was most similar to our model in terms of experimental conditions and the performed task (elbow flexion and extension). An ANN was used to estimate force, where the linear envelopes of the EMG signals were used as inputs. Zhang and Zhang \cite{featureselectionforce2016} used a number of time-domain features extracted from EMG signals and fed into ANN estimators, to estimate grasp force. We chose to compare our model to this model, since their experiment was based on dynamic contractions similar to ours. An SVM with a Gaussian kernel which had been used for force estimation, with rectified-smoothed EMG as the input during both flexion and extension \cite{svrziai2011} and grasping \cite{svrGripforceEST} tasks, was also studied. As well, we considered an SVM with polynomial and linear kernels for further evaluation and comparison. We implemented these methods to the best of our ability based on the descriptions provided in the respective papers. Where certain parameters were not provided, we made the necessary assumptions required to maximize performance.

In order to evaluate the impact of each major component (base learner) in our model, we performed ablation experiments by removing each modality (EMG$_{time}$, EMG$_{freq}$, IMU) from the pipeline. Should the performance drop when a particular input is excluded, it can be concluded that the input in question makes a positive contribution towards the final goal of accurate force prediction.

To evaluate the choice of feature-level fusion used in our deep multimodal CNN, we implemented and analyzed the two additional multimodal strategies, namely input-level fusion and score-level fusion. For input-level fusion, we concatenated the inputs and fed them into a CNN. The architecture of the CNN was designed empirically to obtain the best performance. Two conv-blocks with $64$ and $128$ filters were used (filter size was $3 \times 3$), and the max-pooling was $3 \times 3$. One fully connected layer with $128$ neurons and a regression layer were used. \textit{L}2 regularization was applied to avoid overfitting, and dropout was performed after the final conv-block. For score-level fusion, the same base learners used in our model were utilized. We trained separate base learners and averaged the outcomes to obtain the estimated force.

\subsection{Statistical Analysis}

Statistical analyses are performed to investigate whether the $R^2$ values obtained for the considered conditions were statistically significantly different and whether the proposed deep multimodal CNN had significantly better performance compared to the other configurations. This analysis is performed by comparing the intra-subject and the inter-subject errors across methods using the Friedman test which is a non-parametric statistical test, as our data were not normally distributed. The Friedman is similar to the parametric repeated measures ANOVA, where both of them have been used to detect differences in methods across multiple measurement. The null-hypothesis  is rejected when the p-value is below the critical value, which was set to 0.05. Nemenyi post-hoc testing was applied to perform pairwise multiple comparison tests to determine which pairs of variables have significant differences.

\section{Results and Discussion}

In this section we present the performance of our method and evaluate the impact of different components and parameters of the model as discussed previously in Section \ref{evaluation} as well as the effect of segment size. For the performance of our proposed deep multimodal CNN, the holdout test set results are reported. For the ablation study, fusion strategy, and segment size, where we assess the impact of different components and parameters, results on the validation set are reported as we tuned the components and parameters of our model on these data.

\subsection{Performance}

The results of the proposed deep multimodal CNN, using EMG$_{time}$, EMG$_{freq}$, and IMU, are presented in Table \ref{intra-inter-comp_others}. 
Also, Table \ref{intra-inter-comp_others} shows results of our comparison study with the other force estimation schemes from the literature. Bold values denote the best performance while the asterisk (*) denotes statistically significant difference with respect to our method. In considering other studies, we did not compare our model’s performance with results reported in the literature due to differences in the experimental conditions, participant characteristics, and the types of EMG electrodes and other instrumentation used. As well, the criteria used to assess performance varies between studies. Thus, we re-implemented the proposed methods, as described in \ref{evaluation}, and used our dataset to estimate force and compare the selected approaches to our models.
The statistical analysis indicates that our method has significantly outperformed the other methods in both schemes, except for inter-subject modelling in one case \cite{featureselectionforce2016}. This method \cite{featureselectionforce2016} had the highest $R^2$ values compared to other methods considered; its ANN model used several extracted features from the data, indicating the dependency of ANN models on hand-crafted features. It should also be stated that the methods proposed by the related work \cite{mobasser2012ANN,featureselectionforce2016,svrziai2011,svrGripforceEST}, have originally been designed to meet the particular goals of the papers. For instance, the choice of sensors, muscle groups, contraction types, experimental setup, and other factors may result in lack of sufficient generalization of these methods to our data and experimental conditions. Nonetheless, since these works provide a solution for EMG-based force estimation, a direct comparison will indicate the relative performance of our approach.

\begin{table}[h]
\caption{The $R^2$ values (mean±SD) for the proposed method in comparison with related work, for all experimental conditions in intra- and inter-subject schemes.}
\centering
\label{intra-inter-comp_others}
\scalebox{0.85}{
\centering
\begin{tabular}{llll}
\hline
Method & Isotonic & Isokinetic & Dynamic \\ \hline\hline
\textit{Intra-subject Modelling}\\ \hline
Mobasser et al. \cite{mobasser2012ANN}  & 0.58$\pm$0.220$^{\ast}$ & 0.39$\pm$0.510$^{\ast}$ & 0.38$\pm$0.460$^{\ast}$\\
Zhang and Zhang \cite{featureselectionforce2016} & 0.69$\pm$0.260$^{\ast}$ & 0.52$\pm$0.380$^{\ast}$& 0.46$\pm$0.440$^{\ast}$ \\
Ziai et al. \cite{svrziai2011}, Castellini et al. \cite{svrGripforceEST} & 0.62$\pm$0.210$^{\ast}$& 0.49$\pm$0.260$^{\ast}$& 0.45$\pm$0.220$^{\ast}$\\
SVM (Polynomial)& 0.54$\pm$0.310$^{\ast}$& 0.41$\pm$0.550$^{\ast}$& 0.32$\pm$0.430$^{\ast}$\\
SVM (Linear)& 0.51$\pm$0.140$^{\ast}$& 0.33$\pm$0.390$^{\ast}$& 0.28$\pm$0.510$^{\ast}$\\
\textbf{Deep multimodal CNN}  &   \textbf{0.91}$\pm$\textbf{0.034} &\textbf{0.87}$\pm$\textbf{0.041} & \textbf{0.81}$\pm$\textbf{0.037} \\
\hline\hline
\textit{Inter-subject Modelling}\\ \hline
Mobasser et al. \cite{mobasser2012ANN}  & 0.49$\pm$0.026$^{\ast}$ & 0.36$\pm$0.050$^{\ast}$ & 0.36$\pm$0.048$^{\ast}$\\
Zhang and Zhang \cite{featureselectionforce2016} & 0.52$\pm$0.031 & 0.41$\pm$0.083& 0.40$\pm$0.083 \\
Ziai et al. \cite{svrziai2011}, Castellini et al.\cite{svrGripforceEST} & 0.44$\pm$0.061$^{\ast}$ & 0.36$\pm$0.052$^{\ast}$ & 0.38$\pm$0.061$^{\ast}$\\
SVM (Polynomial) & 0.42$\pm$0.056$^{\ast}$& 0.28$\pm$0.033$^{\ast}$& 0.21$\pm$0.033$^{\ast}$\\
SVM (Linear)& 0.36$\pm$0.051$^{\ast}$& 0.26$\pm$0.090$^{\ast}$& 0.16$\pm$0.038$^{\ast}$\\
\textbf{Deep multimodal CNN} & \textbf{0.81}$\pm$\textbf{0.048}  &  \textbf{0.64}$\pm$\textbf{0.037} & \textbf{0.59}$\pm$\textbf{0.042} \\ 
\hline
\end{tabular}}
\end{table}

\begin{figure}
    \centering
    \includegraphics[width=0.90\columnwidth]{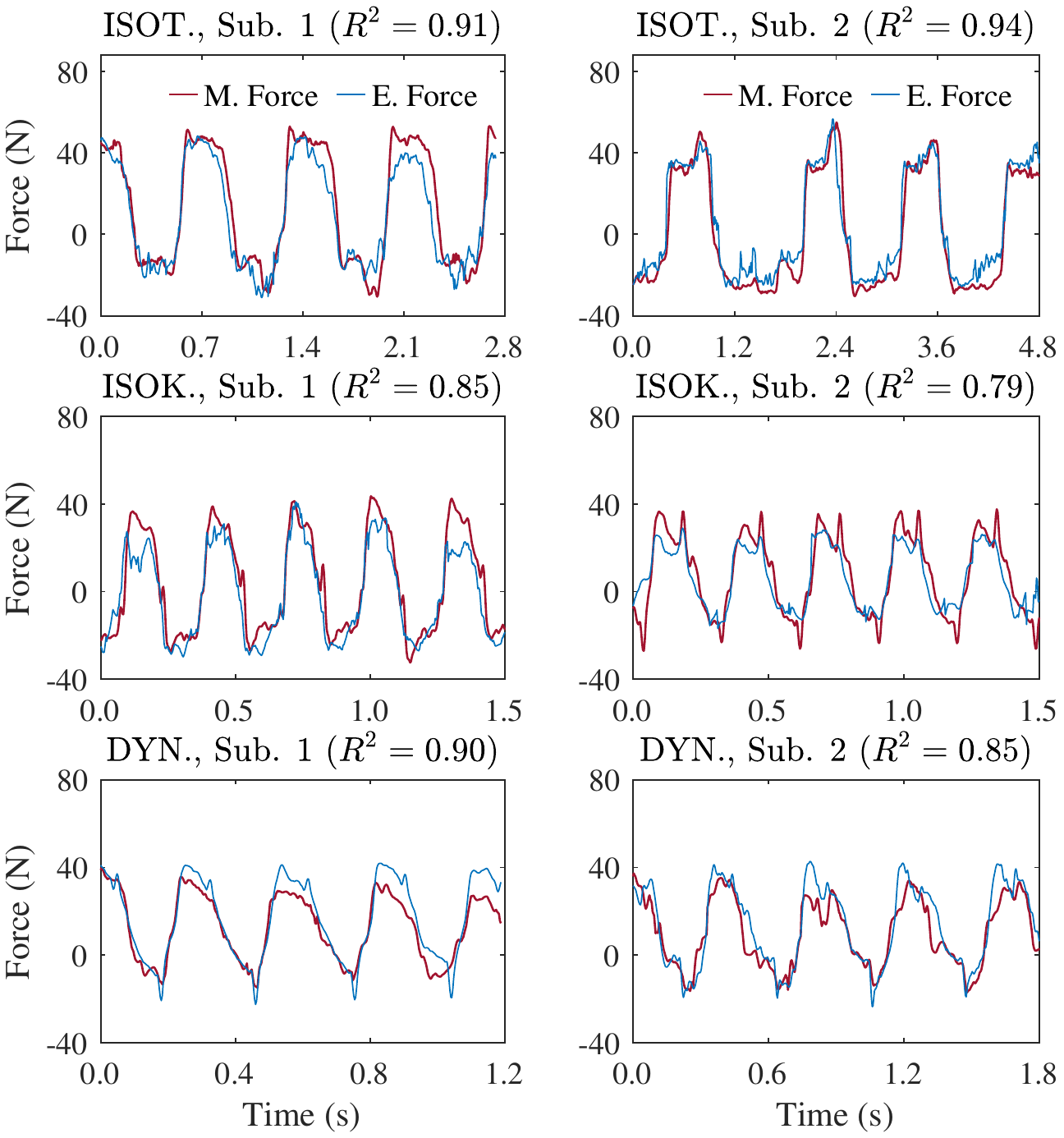}
    \caption{Estimated (E.) force versus measured (M.) force for two sample subjects under different experimental conditions, isotonic (ISOT.), isokinetic (ISOK.), and  dynamic (DYN.) contractions, on the holdout test sets.}
    \label{fig_hold1}
\end{figure}

The force estimation results for the intra-subject scheme (all three experimental conditions) are  accurate, as the $R^2$ values are 
high and small standard deviations are obtained.  Fig. \ref{fig_hold1} illustrates the measured and estimated force values for the holdout test sets of two sample subjects.

The inter-subject performance is expectedly lower than intra-subject, despite the use of deeper networks. This difference is mainly due to the physiological and non-physiological differences among participants which make the inter-subject force modelling more complex. Additionally, we observe that the isotonic, inter-subject results show higher $R^2$ values compared to the isokinetic and dynamic results.
As per the nature of isokinetic experimental conditions, the velocity is consistent (both in flexion and extension), whereas the velocity is more variable in the isotonic and dynamic cases, given the lack of control on this parameter under these conditions. Moreover, 
the force is relatively consistent over time and between the subjects in the isotonic case, where it is not consistent in the isokinetic and dynamic cases.
It seems that the variability in force level is a greater contributing factor to the increased error in inter-subject modeling under isokinetic and dynamic conditions, than the variability in velocity. The reason for this could be due to the incorporation of the IMU data which makes the model more robust to changes in position and velocity.

\subsection{Ablation Study} \label{Ablation}

Table \ref{intra-inter-ablation} shows the $R^2$ values for each modality, illustrating its impact on the model performance for intra- and inter-subject schemes, under different experimental conditions. As discussed earlier, the results on the validation set are reported, as the effects of different components on our model are investigated.

For the intra-subject scheme, the model using EMG$_{time}$ and EMG$_{freq}$ shows significant differences for all conditions versus the proposed deep multimodal CNN, indicating the necessity for including the IMU data. The exclusion of EMG$_{freq}$ shows insignificant differences for all three conditions. The exclusion of EMG$_{time}$ results in significant deterioration of performance under isokinetic and dynamic conditions.

For the more challenging inter-subject scenario, we observe that IMU and EMG$_{time}$ are both necessary for the model as their individual exclusions result in significant reduction of $R^2$ values. EMG$_{freq}$, however, does show significant importance for the isotonic scenario, while the reductions in accuracy in isokinetic and dynamic conditions are not significant.

It can be observed that while there are clear advantages for including both EMG$_{time}$ and IMU for force estimation, the use of EMG$_{freq}$ has limited contribution towards force estimation with insignificant improvement for most cases except for inter-subject modeling of isotonic contractions. Nonetheless, given the availability of the EMG$_{freq}$ and its simple addition as part of the pipeline, it can be argued that its inclusion can result in more robustness and generalizability across different experimental conditions (e.g. isotonic).

\begin{table}[t]
\caption{The $R^2$ values (mean±SD) for the ablation study between different models for isotonic, isokinetic, and dynamic conditions for intra- and inter-subject schemes, using the validation set.}
\label{intra-inter-ablation}
\centering
\scalebox{0.90}{
\begin{tabular}{llll}
\hline
Method & Isotonic & Isokinetic & Dynamic \\ \hline\hline
\textit{Intra-subject Modelling}\\ \hline
\textbf{Deep multimodal CNN}    &  \textbf{0.98}$\pm$\textbf{0.002} & \textbf{0.96}$\pm$\textbf{0.008} & \textbf{0.96}$\pm$\textbf{0.004} \\
EMG$_{time}$, EMG$_{freq}$  & 0.86$\pm$0.041$^{\ast}$ & 0.44$\pm$0.330$^{\ast}$ & 0.64$\pm$0.039$^{\ast}$\\
EMG$_{time}$, IMU & 0.98$\pm$0.002 & 0.94$\pm$0.008& 0.96$\pm$0.004 \\
EMG$_{freq}$, IMU & 0.96$\pm$0.018& 0.89$\pm$0.210$^{\ast}$& 0.90$\pm$0.007$^{\ast}$\\
\hline\hline
\textit{Inter-subject Modelling}\\ \hline
\textbf{Deep multimodal CNN}   & \textbf{0.89}$\pm$\textbf{0.040}  &  \textbf{0.68}$\pm$\textbf{0.022} & \textbf{0.69}$\pm$\textbf{0.016}\\
EMG$_{time}$, EMG$_{freq}$  &0.69$\pm$0.014$^{\ast}$ & 0.40$\pm$0.110$^{\ast}$& 0.29$\pm$0.081$^{\ast}$\\
EMG$_{time}$, IMU & 0.81$\pm$0.015$^{\ast}$& 0.66$\pm$0.018 & 0.68$\pm$0.021 \\
EMG$_{freq}$, IMU & 0.73$\pm$0.008$^{\ast}$ & 0.61$\pm$0.013$^{\ast}$& 0.63$\pm$0.036$^{\ast}$ \\
\hline
\end{tabular}}
\end{table}

Using the EMG signal only (in both domains) for the isotonic case can provide acceptable accuracy for force estimation, despite the changing velocity, compared to other conditions. However, even during the isotonic contraction, using our method (deep multimodal CNN), which concatenates the IMU features with EMG$_{time}$ and EMG$_{freq}$, improved the performance by $13.95\%$ and $28.98\%$
for intra- and inter-subject schemes, respectively. Adding kinematic information resulted in better improvement in the force modelling performance for the isokinetic and dynamic cases. This could be because the IMU tracks changes in the elbow joint angle, which is related to the muscle length and velocity, and provides information regarding the force-length and force-velocity properties of the muscles.

\subsection{Impact of Fusion Strategy} \label{Fusion Strategy impact}

The results of the comparison of feature-level, input-level and score-level fusion are presented in Table \ref{Fusion Strategy_intra_inter} for intra- and inter-subject schemes. For all experimental conditions, feature-level fusion outperforms the other fusion strategies, for both schemes.

\begin{table}[t]
\caption{The $R^2$ values (mean±SD) for comparing the fusion strategies between different models for isotonic, isokinetic, and dynamic contractions for intra- and inter-subject schemes, using the validation set.}
\label{Fusion Strategy_intra_inter}
\centering
 \scalebox{0.90}{
\begin{tabular}{llll}
\hline
Method & Isotonic & Isokinetic & Dynamic \\ \hline\hline
\textit{Intra-subject Modelling}\\ \hline
Input-level Fusion & 0.65$\pm$0.028$^{\ast}$ & 0.56$\pm$ 0.068$^{\ast}$& 0.55$\pm$0.081$^{\ast}$\\
Score-level Fusion & 0.58$\pm$0.360$^{\ast}$ & 0.46$\pm$0.440$^{\ast}$& 0.56$\pm$0.490$^{\ast}$\\
\textbf{Deep multimodal CNN} &\textbf{0.98}$\pm$\textbf{0.002}  & \textbf{0.96}$\pm$\textbf{0.008} & \textbf{0.96}$\pm$\textbf{0.004}\\
\hline\hline
\textit{Inter-subject Modelling}\\ \hline
Input-level Fusion & 0.52$\pm$0.018$^{\ast}$ & 0.20$\pm$ 0.092$^{\ast}$& 0.36$\pm$0.048$^{\ast}$\\
Score-level Fusion & 0.44$\pm$0.010$^{\ast}$& 0.31$\pm$0.040$^{\ast}$& 0.27$\pm$0.002$^{\ast}$ \\
\textbf{Deep multimodal CNN} & \textbf{0.89}$\pm$\textbf{0.040}  &  \textbf{0.68}$\pm$\textbf{0.022} & \textbf{0.69}$\pm$\textbf{0.016} \\
\hline
\end{tabular}}
\end{table}

\subsection{Impact of Model Parameters}

\subsubsection{Number of Convolutional Blocks}

\begin{figure*}
    \centering
    \includegraphics[width=14.5 cm]{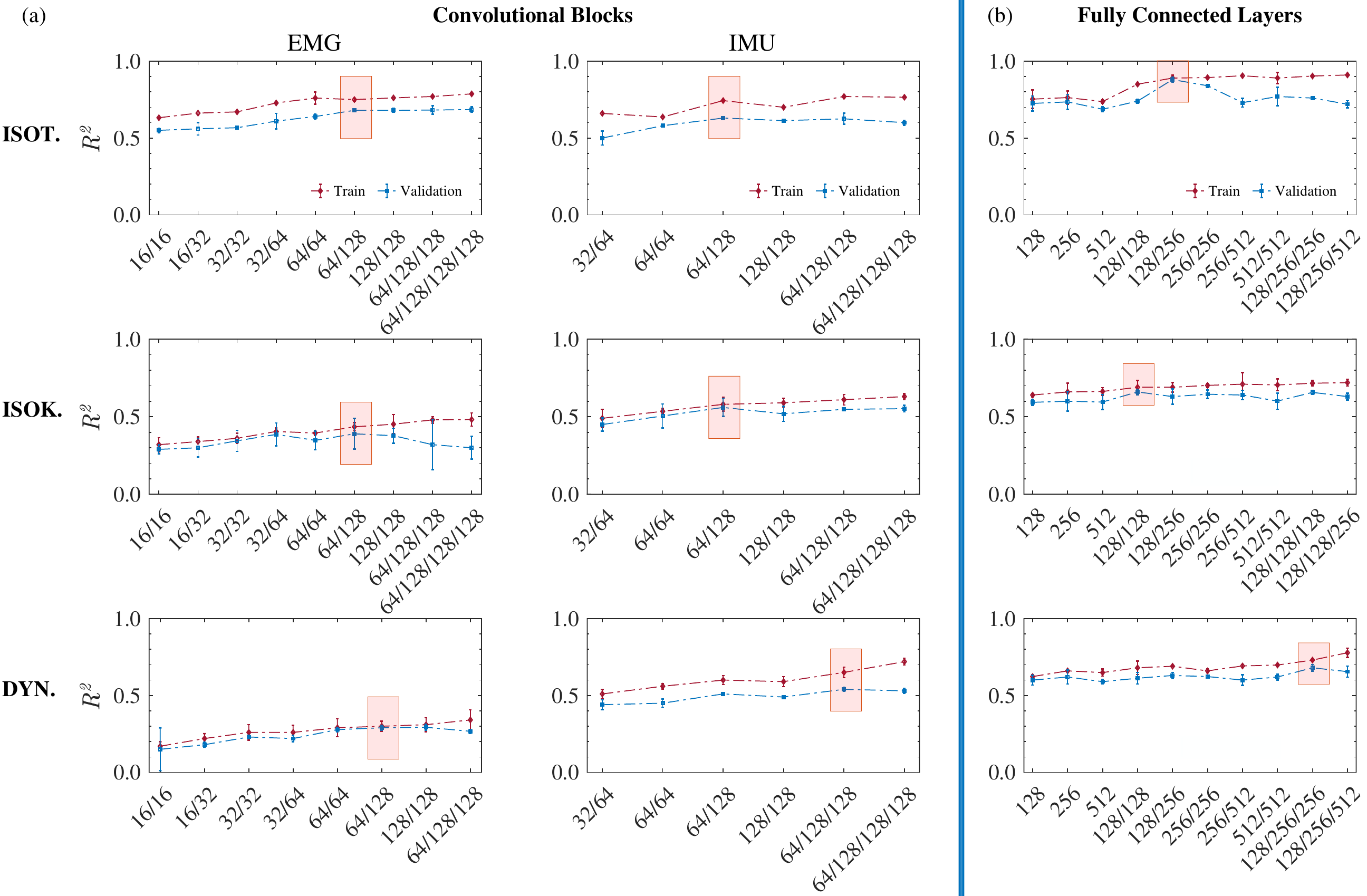}
    \caption{Inter-subject models developed for different numbers of (a) convolutional blocks (EMG and IMU learners), and (b) fully connected layers and neurons, during isotonic (ISOT.), isokinetic (ISOK.), and  dynamic (DYN.) contractions.}
    \label{inter_base_Learner_fc}
\end{figure*}

As discussed earlier in Section \ref{convblocksIntra} and \ref {convblocksInter}, for intra-subject force modelling, a relatively shallow model was sufficient for accurate estimation of the force, while for the inter-subject scheme, deeper models were required given the complexity, non-linearities, and physiological differences across different participants. Fig. \ref{inter_base_Learner_fc}(a) presents the results of our experiments on different numbers of layers and layer sizes for the base learners for the inter-subject scheme. The selected models in each case are shown within the pink boxes, where the selected models had higher $R^2$ for the validation set compared to other configurations. It can be seen that in most cases, the training performance is enhanced as the models become more complex, while we generally observe a point of diminishing improvements on the validation set beyond a certain point. In such cases, we prioritized the validation set performance to avoid overfitting.

Given that based on our ablation tests  the EMG$_{freq}$ showed rather poor performance when used alone for inter-subject modelling, we opted to not perform an individual search and simply utilize the same architecture as the EMG$_{time}$. We should re-iterate, however, that EMG$_{freq}$ did contribute to the pipeline in some cases, and even resulted in significant improvement for inter-subject modeling of isotonic contractions.

\subsubsection{Number of Fully Connected Layers}
We examined different numbers of fully connected layers and neurons for inter-subject modelling, as shown in Fig. \ref{inter_base_Learner_fc}(b).

For the isotonic case two fully connected layers with $128$ and $256$ neurons, and for the isokinetic case, two layers with $128$ and $128$ neurons, results in the best performance. For the dynamic case, three fully connected layers with $128$, $256$, and $256$ neurons were required. This could be due to the more complex nature of force modelling during this type of contraction compared to the quasi-dynamic cases. One fully connected layer with one neuron, which acts as a regression layer, is used at the end (for all experimental conditions) to estimate the generated force. 

Additionally, comparing the multimodal CNN approach (Fig. \ref{inter_base_Learner_fc}(b)) to the base learners (Fig. \ref{inter_base_Learner_fc}(a)) for each experimental condition indicates that concatenating the extracted features from different signal modalities results in improvements in force modelling.

\subsubsection{Segment Length} \label{windowsize}

The effect of segment length on the performance under the three conditions was investigated, where segment lengths of $50$, $100$, and $150$ ms were considered to select the optimum window size for force modelling. We evaluated the performance of the model for the intra-subject scheme, where the average of $R^2$ values across subjects are presented in Table \ref{window duration}. The highest performance
was achieved when the segment size was $50$ ms for the deep multimodal CNN, for all experimental conditions. The statistical analysis showed that $50$ ms window size is significantly better than other considered window sizes for isokinetic and dynamic contractions. 
However, for isotonic contractions, $50$ ms is not significantly different than $100$ ms. A reason for this observation could be the variations in the EMG signal and force during isokinetic and dynamic contractions
compared to the isotonic case, in which the EMG amplitude is approximately constant (it follows a constant torque level). The $50$ ms window showed significant improvement over $150$ ms in every scenario. These findings are in contrast with those of Farrell et al. \cite{farrell2007optimal} in terms of optimal window size, as they observed improvements from $50$ ms to $100$ ms, although this improvement was not significant. However, they considered isometric contractions, which could account for the difference in results. Thus, in order to estimate force more accurately, the segment length should be selected so that the EMG is almost stationary throughout the duration of the segment. Therefore, the segment size was set to $50$ ms for data segmentation for the analysis of the quasi-dynamic and dynamic cases.

\begin{table}[t]
\caption{The $R^2$ values (mean±SD) for different segment length, under different experimental conditions, for intra-subject scheme, for the validation set.}
\label{window duration}
\centering
\scalebox{0.95}{
\begin{tabular}{lccc}
\hline
Conditions    & $50$ ms       & $100$ ms     & $150$ ms    \\ \hline\hline
Isotonic & \textbf{0.98}$\pm$\textbf{0.002} & 0.97$\pm$0.010  & 0.92$\pm$0.017$^{\ast}$ \\
Isokinetic & \textbf{0.96}$\pm$\textbf{0.008} & 0.90$\pm$0.049$^{\ast}$  & 0.83$\pm$0.063$^{\ast}$ \\
Dynamic &\textbf{0.96}$\pm$\textbf{0.004} & 0.92$\pm$0.026$^{\ast}$ & 0.82$\pm$0.037$^{\ast}$ \\
\hline
\end{tabular}}
\end{table}

\vspace{10pt}

\subsection{Impact of IMU Sensors}

The IMU used in this study is comprised of three sensors, namely a triaxial accelerometer, gyroscope, and magnetometer. We investigate the impact of each of these sensors on force modelling for different experimental conditions, and both intra- and inter-subject schemes. The $R^2$ values (mean±SD) obtained using the IMU are compared to using each individual sensor under the different experimental conditions, for 50 ms segment lengths, are shown in Table \ref{imustudy}.

Our results indicate that using all IMU data for force modelling outperforms each individual sensor (accelerometer, gyroscope, and magnetometer), for all experimental conditions and both schemes. There were no significant differences among the modelling results of the individual sensors. Additionally, although using IMU data alone did not provide good force modelling performance, our results from Section \ref{Ablation} indicate that incorporating IMU data with EMG signal considerably contributes to force estimation. Thus, the IMU data are an important source of information to enhance the force modelling performance  under dynamic conditions, when all sensors are considered.

\begin{table}[t]
\caption{The $R^2$ values (mean±SD) for the IMU and its individual sensors for different experimental conditions (isotonic, isokinetic, and dynamic), and for intra- and inter-subject schemes, using the holdout set. Acc. stands for accelerometer, Gyro. for gyroscope, and Mag. for magnetometer.}
\label{imustudy}
\centering
\scalebox{0.90}{
\begin{tabular}{llll}
\hline
Method & Isotonic & Isokinetic & Dynamic \\ \hline\hline
\textit{Intra-subject Modelling}\\ \hline
\textbf{IMU}&  $\textbf{0.44}\pm\textbf{0.18}^{\ast}$  & $\textbf{0.66}\pm\textbf{0.12}^{\ast}$  & $\textbf{0.48}\pm\textbf{0.16}^{\ast}$  \\
Acc.  & $0.22\pm0.29$ & $0.52\pm0.14$ & $0.22\pm0.21$\\
Gyro. & $0.21\pm0.34$ & $0.43\pm0.14$& $0.23\pm0.34$ \\
Mag. & $0.31\pm0.18$& $0.41\pm0.12^{\ast}$& $0.23\pm0.27^{\ast}$\\
\hline\hline
\textit{Inter-subject Modelling}\\ \hline
\textbf{IMU}   & $\textbf{0.51}\pm\textbf{0.03}^{\ast}$  &  $\textbf{0.53}\pm\textbf{0.03}^{\ast}$ & $\textbf{0.45}\pm\textbf{0.02}^{\ast}$\\
Acc. &$0.39\pm0.08$& $0.25\pm0.15$& $0.32\pm0.12$\\
Gyro. & $0.22\pm0.05^{\ast}$& $0.32\pm0.08$ & $0.24\pm0.09$ \\
Mag. & $0.39\pm0.17$ & $0.33\pm0.13$& $0.27\pm0.06$ \\
\hline
\end{tabular}}
\end{table}

\vspace{10pt}

\section{Conclusion}
Four HD-EMG arrays were used to record EMG signals from the elbow flexor and extensor muscles during isotonic, isokinetic, and fully dynamic elbow flexion and extension. The purpose of this study was to estimate the generated force at the wrist accurately for all three conditions in intra- and inter- subject manners. The proposed  method, deep multimodal CNN, extracted features from EMG signals in different domains and IMU data, using CNN models. Then, the extracted features were concatenated and fed into dense layers to be weighted for the force estimation. We obtained accurate results for both intra- and inter-subject schemes, for all experimental conditions, where our solutions outperformed a number of previously published methods. The ablation experiments showed that force estimation improved significantly when the kinematic information was considered. Lastly, we explored two other fusion strategies, input-level and score-level fusions, where the results confirmed the effectiveness of our proposed method than other fusion strategies. Thus, our proposed method can be of large benefit for prosthesis control and assistive devices.

In future work, a recurrent neural network will be used for the force estimation instead of the shallow neural network, with the features extracted using CNN models. Additionally, anthropometric information from participants such as forearm and upper arm length as well as their circumference can  be considered to improve the inter-subject modelling performance. Finally, it should be noted that all the recruited participants in our study were healthy adults. Participants with neuromuscular disorders could be considered to investigate the performance of the developed models for force estimation in weakened or debilitated muscles. 

\bibliographystyle{IEEEtran}  
\bibliography{Gelare}

\end{document}